\documentclass[11pt,twoside]{article}

\usepackage{asp2021}

\aspSuppressVolSlug
\resetcounters

\begin{document}

\title{\texttt{tilepy}: A Flexible Open-Source Scheduling Engine for Time-Domain and Multi-Messenger Astronomy}

\author{Fabian Schussler,$^1$ H. Ashkar,$^2$ W. Kiendrébéogo,$^1$ M. Seglar-Arroyo,$^3$ M. de Bony$^{1,4}$, A. Berti$^5$ E. Ruiz-Velasco$^6$, and R. Le Montagner$^7$}
\affil{$^1$IRFU, CEA, Université Paris-Saclay, France; \email{fabian.schussler@cea.fr}} \quad \affil{$^2$ICE-CSIC, Barcelona, Spain \: $^3$IFAE, Barcelona, Spain \: $^4$CPPM, Marseille, France \: $^5$MPP, Munich, Germany \: $^6$LAPP, Annecy, France \: $^7$IJCLab, Orsay, France}

\paperauthor{Fabian Schussler}{fabian.schussler@cea.fr}{0000-0003-1500-6571}{Université Paris-Saclay}{IRFU}{Paris-Saclay}{}{}{France}



\begin{abstract}
The era of multi-messenger astrophysics requires rapid and efficient follow-up of transient events, many of which, such as gravitational waves (GW), gamma-ray bursts (GRB), and high-energy neutrinos, suffer from poor sky localisation. We present \texttt{tilepy}, a Python-based software designed to optimize observation schedules for these events. We here detail the modular architecture of \texttt{tilepy}, which separates high-level scheduling logic from low-level tiling and pointing tools, enabling full adaptability for ground- and space-based observatories. Furthermore, we describe the integration of \texttt{tilepy} into the Astro-COLIBRI platform, providing the community with a user-friendly interface and API for triggering complex observation campaigns in real time.
\end{abstract}

\vspace{-1cm}
\section{Introduction}
Time-domain astrophysics has been transformed by the direct detection of gravitational waves and high-energy neutrinos. However, a significant challenge remains: the localization uncertainty of these transients often spans tens to thousands of square degrees~\citep{2023PhRvX..13d1039A, 2025ApJ...995L..18A}. Since most electromagnetic follow-up facilities possess fields of view (FoV) significantly smaller than these error boxes, efficient tiling strategies are essential to maximize the probability of detecting a counterpart.

\texttt{tilepy} was developed to address this optimization problem. It maximizes the probability of counterpart detection by leveraging both, 2D skymaps as well as 3D galaxy catalogs, while strictly adhering to observatory constraints such as visibility, background light conditions, and instrument limits \citep{2024ApJS..274....1S}. It is available under the GNU Lesser General Public License via GitHub~\citep{tilepy}.

While previous works have demonstrated \texttt{tilepy}'s application to specific campaigns like GW170817~\citep{2017ApJ...850L..22A} and other GW follow-up observations~\citep{2021ApJ...923..109A}, we here focus on the software's internal architecture and its integration into the broader multi-messenger ecosystem via Astro-COLIBRI.

\section{Software Architecture}

\begin{figure}
    \centering
    \includegraphics[width=0.75\linewidth]{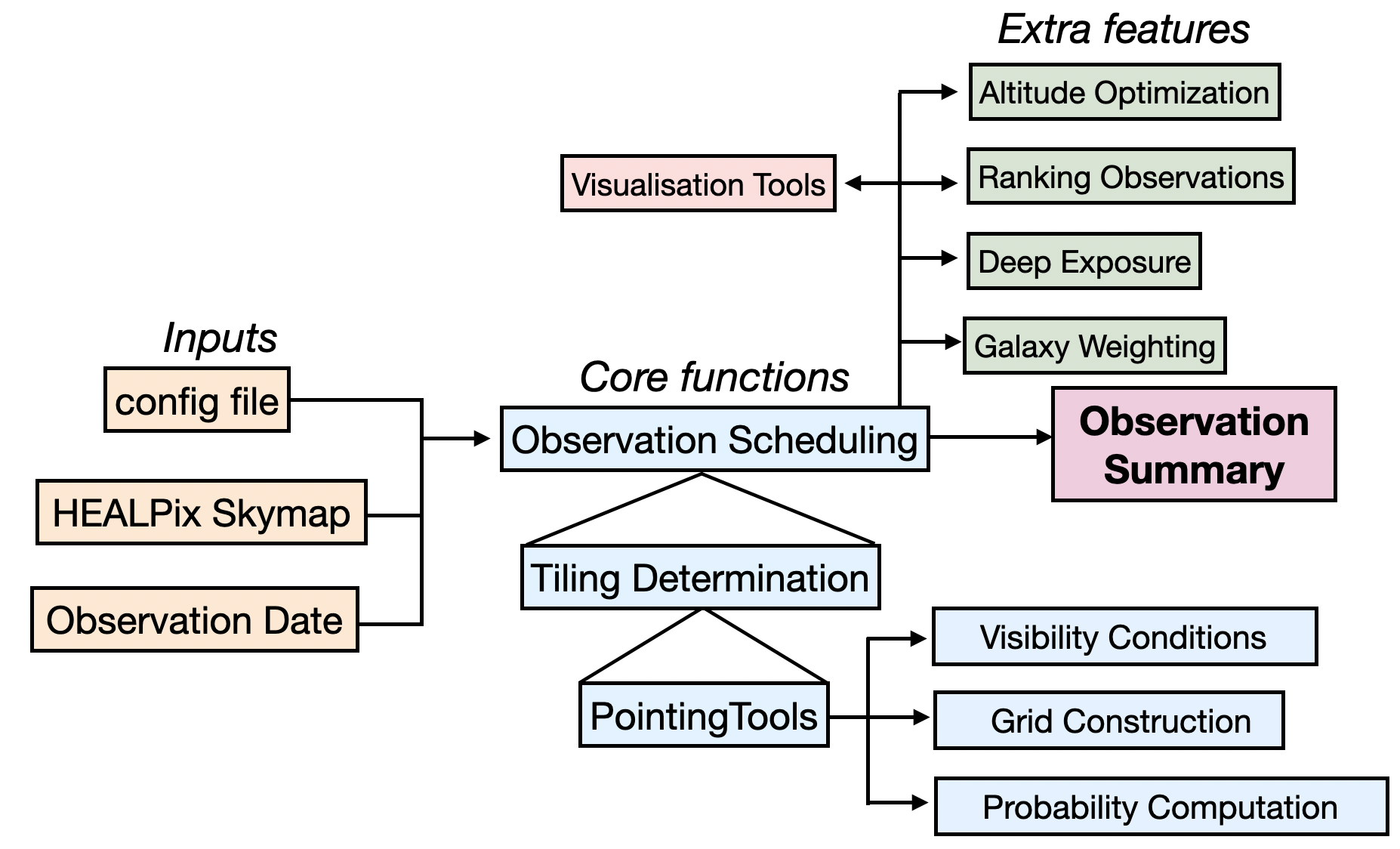}
    \caption{Outline of the \texttt{tilepy} architecture. From~\cite{2024ApJS..274....1S}}
    \label{fig:arch}
\end{figure}

The design of \texttt{tilepy} follows a modular, hierarchical structure. This ensures maintainability and allows users to customize strategies without altering the core scheduling logic. The architecture is illustrated in Figure \ref{fig:arch} and can be categorized into three primary layers: Inputs, Core Functions, and Outputs.

\subsection{Inputs and Configuration}
The software requires three primary inputs to initialize a campaign. First, a localization probability map of the transient (e.g., from LIGO/Virgo/KAGRA, Fermi-GBM, or IceCube). Typically given in {\it HEALPIX} format. Second, the proposed start time for the campaign and finally, a distinct configuration file is required for each participating telescope. These files define the observatory's location, FoV size and shape, horizon limits, and specific constraints (e.g., Moon avoidance angles). This modularity allows \texttt{tilepy} to schedule multi-observatory campaigns simultaneously by simply loading multiple configuration files.

\subsection{Core Functions}
The processing logic is divided into three levels of abstraction, moving from global, high-level scheduling to low-level computation:

\subsubsection{Observation Scheduler}
This is the top-level manager. It parses the configuration files and the input skymap. It determines the appropriate strategy (e.g., 2D probability integration vs. 3D galaxy targeting) based on the available data (e.g., whether the event includes distance information) and user requests. By default, the scheduler employs a ``greedy'' approach: it identifies the highest-probability region available at the earliest possible time slot, schedules it, masks that region to prevent duplicate observations, and iterates until the observing window closes or probability thresholds are met. Details are given in~\citep{2021JCAP...03..045A}.

\subsubsection{Tiling Determination}
The tiling layer handles the geometry of the search. It supports varying grid resolutions, balancing computational speed with accuracy. Users can specify the required resolution (via the HEALPix \texttt{NSIDE} parameter) or allow the software to optimize it based on the telescope's FoV. Recent updates have expanded this layer to support complex FoV shapes. Beyond standard circular fields, \texttt{tilepy} now supports polygonal shapes (squares, hexagons) and arbitrary rotations, which is critical for accurate modeling of instrument footprints like those of the Vera Rubin Observatory, Swift-XRT, etc.

\subsubsection{Pointing Tools}
This low-level layer performs the mathematical heavy lifting. Its core components include \textit{Visibility Handling}, which computes the altitude/azimuth of targets over time, checking against constraints (Sun/Moon altitude, Earth limb for satellites). The \textit{Probability Computation} integrates probability density over the defined FoV. For 3D strategies, it queries galaxy catalogs (e.g., GLADE+~\citep{2022MNRAS.514.1403D}) to weigh pointings by stellar mass or galaxy luminosity. New modules supporting \textit{Space-Based Observatories} have been added recently to calculate spacecraft trajectories, identifying occultations by the Earth, Sun, and Moon, and flagging passages through the South Atlantic Anomaly (SAA).

\subsection{Outputs and Visualization}
The final layer generates the observation schedule (coordinates and timing) and a suite of visualization tools. These include coverage maps, visibility plots (altitude vs. time), and probability accumulation curves, aiding observers in assessing the quality of the schedule.

\section{API and integration with Astro-COLIBRI}

To make advanced scheduling accessible to the wider community, \texttt{tilepy} has been integrated into the Astro-COLIBRI platform~\citep{2021ApJS..256....5R, 2023Galax..11...22R}. This integration bridges the gap between alert reception and observation planning.

The core functionality of \texttt{tilepy} is hosted as a public API~\footnote{\url{https://tilepy.com}}. This decoupling allows external services to request schedules without installing the software or its dependencies locally. The API accepts an event identifier (or a skymap URL) and observatory parameters, and returns a JSON-formatted schedule. This architecture supports the ``Trigger'' model, in which automatic transient handlers can programmatically request observation plans immediately upon receiving an alert.

For manual users, \texttt{tilepy} is embedded directly into the Astro-COLIBRI web and mobile interfaces. Users viewing an event in Astro-COLIBRI can generate an optimized tiling plan with a single click. This integration democratizes access to complex scheduling algorithms and allows citizen scientists to visualize the visibility of a GW event and obtain a mathematically optimized pointing list without writing code. The platform also hosts a dedicated helpdesk and discussion forum\footnote{\url{https://forum.astro-colibri.science/c/instrumentation-and-tools/tilepy}} to support \texttt{tilepy} users.

\section{Status and Future Outlook}
The \texttt{tilepy} ecosystem is fully operational and utilized by major collaborations like H.E.S.S., and the first CTAO Large-Sized Telescope (LST-1). The connection to Astro-COLIBRI and the public API ensure that the optimized observation schedule are widely available and easily accessible also by citizen scientists and amateur astronomers.

\texttt{tilepy} provides a robust, modular solution for the complex problem of multi-messenger follow-up. Its architecture, separating configuration from core logic, allows it to serve a diverse range of facilities, from ground-based Cherenkov telescopes to space-based X-ray observatories. The integration with Astro-COLIBRI represents a significant step forward in usability, allowing real-time access to high-performance scheduling tools. Future developments will focus on further optimizing algorithms for "Travelling Salesman" style slew-minimization (possibly using AI aided approaches), expanding support for non-HEALPix localization formats.

\section*{Acknowledgments}
We acknowledges the support of the French Agence Nationale de la Recherche (ANR) via the project "Multi-messenger Observations of the Transient Sky (MOTS)" under reference ANR-22-CE31-0012. We also acknowledge support from the "Astrophysics Centre for Multi-messenger studies in Europe (ACME)" funded by the European Union’s Horizon Europe Research and Innovation programme under grant no. 101131928. 

\bibliography{117}  


\end{document}